\documentclass[prb,showpacs,superscriptaddress,twocolumn,notitlepage]{revtex4-1}

\usepackage{graphicx, subfigure}

\usepackage[utf8x]{inputenc} 

\usepackage{amsmath}
\usepackage{amsfonts}
\usepackage{graphicx}

\usepackage{natbib}
\usepackage[final=true]{hyperref}

\usepackage{color}
\usepackage[usenames,dvipsnames]{xcolor}

\begin{document}
\title{Novel approach to Raman spectra of nanoparticles }

\author{S.~V.~Koniakhin}
\email{kon@mail.ioffe.ru}
\affiliation{Institut Pascal, PHOTON-N2, University Clermont Auvergne, CNRS, 4 avenue Blaise Pascal, 63178 Aubi\`{e}re Cedex, France}
\affiliation{St. Petersburg Academic University - Nanotechnology Research and Education Centre of the Russian Academy of Sciences, 194021 St. Petersburg, Russia}
\affiliation{Ioffe Physical-Technical Institute of the Russian Academy of Sciences, 194021 St.~Petersburg, Russia}

\author{O.~I.~Utesov}

\affiliation{Petersburg Nuclear Physics Institute NRC ``Kurchatov Institute'', Gatchina, 188300  St.\ Petersburg, Russia}
\affiliation{St. Petersburg Academic University - Nanotechnology Research and Education Centre of the Russian Academy of Sciences, 194021 St. Petersburg, Russia}

\author{I.~N.~Terterov}

\affiliation{St. Petersburg Academic University - Nanotechnology Research and Education Centre of the Russian Academy of Sciences, 194021 St. Petersburg, Russia}

\author{A.~V.~Siklitskaya}
\affiliation{Institute of Physical Chemistry, Polish Academy of Sciences, 44/52 Kasprzaka 01-224 Warsaw}

\author{D.~Solnyshkov}
\affiliation{Institut Pascal, PHOTON-N2, University Clermont Auvergne, CNRS, 4 avenue Blaise Pascal, 63178 Aubi\`{e}re Cedex, France}

\author{A~.G~.~Yashenkin}
\affiliation{Petersburg Nuclear Physics Institute NRC ``Kurchatov Institute'', Gatchina, 188300  St.\ Petersburg, Russia}
\affiliation{Department of Physics, Saint Petersburg State University, 7/9 Universitetskaya nab., 199034 St. Petersburg, Russia}

\begin{abstract}

In crystalline nanoparticles the Raman peak is downshifted with respect to the bulk material and has asymmetric broadening. These effects are straightly related to the finite size of nanoparticles, giving the perspective to use the Raman spectroscopy as the size probe. By combining the dynamical matrix method (DMM) and the bond polarization model (BPM), we develop a new (DMM-BPM) approach to the description of Raman spectra for random arrays of nanoparticles. The numerical variant of this approach is suitable for the description of small particles, whereas its simplier to implement analytical version allows to obtain the Raman spectra of arbitrary sized particles. Focusing on nanodiamond powders, the DMM-BPM theory is shown to fit the most recent experimental data much better than the commonly used phonon confinement model (PCM), especially for small enough nanoparticles. 

\end{abstract}

\maketitle

\section{Introduction}

Crystalline nanoparticles, including semiconductor particles and nanodiamonds, are actively investigated nowadays for applications in novel materials \cite{mochalin2012properties}, quantum computing \cite{veldhorst2014addressable,Chen2017}, biology and medicine\cite{park2009biodegradable,kim2016ultrasmall}. The outstanding progress in nanoparticle manufacturing techniques dictates the further efforts in characterization and standardization of their size, shape, phase composition, and surface morphology. For such purposes, the high resolution transmission electron microscopy, X-ray diffraction, dynamical light scattering, atomic force microscopy, and other methods are used.

Among others, significant role is played by the Raman spectroscopy. It is the unique long-wavelength (optical) nondestructive experimental
probe that provides the tool to investigate the excitations (phonons, excitons \cite{wang2015double}, magnons \cite{devereaux2007inelastic}, etc.) on the scale of several interatomic distances. This allows to characterize the nanoparticle arrays based on the analysis of the positions of the Raman peaks and their (asymmetric) broadening.

According to the Heisenberg indeterminacy principle, the momentum conservation law in nanoparticles is violated due to localization of photon-phonon interaction within the volume of a particle. While in bulk crystals only the phonons with wavevector $q=0$ contribute to the Raman spectra (RS), in nanoparticles $q$ is quantized due to the size quantization effect. The minimal phonon wavevector is $q_{L} \sim 2\pi/L$, with $L$ being the typical nanocrystallite size. This yields measurable Raman peak downshift, as compared to the bulk material.

The standard method of theoretical analysis of the crystalline nanoparticle RS  \cite{yoshikawa1993raman,yoshikawa1995raman,osswald2009phonon,gao2016size,bahariqushchi2017correlation,arora2007raman} is the semiphenomenological phonon confinement model (PCM) introduced by Richter, Wang, and Ley \cite{richter1981one} and further developed by Campbell and Fauchet \cite{campbell1986effects}. This model is not free of disadvantages. First, PCM is based on the assumption of a smooth (Gaussian) decay of atomic vibration amplitudes from nanocrystallite center to its surface, which seems ill-founded. Indeed, this assumption brings to the theory the adjustable parameter of phonon amplitude at the particle boundary. In our opinion, it is an attempt to incorporate several physical phenomena (such as particle size, shape, disorder, etc.) using a  single quantity. As a result, for the commonly used value of this parameter, 98\% of vibration energy is unphysically restricted to 2\% of nanocrystallite volume near its center \cite{zi1997comparison}.  

Second, within the PCM the RS of identical nanoparticles are broad and smooth. However, as it can be seen from \textit{ab initio} calculations of nanodiamond RS (see Refs. \cite{zhang2005signature,filik2006raman,li2010convergence}) and experimental fullerene RS\cite{chase1992vibrational}, the single nanocrystallite Raman fingerprint is actually very sparse. It has a comb-like shape with multiple narrow peaks. The approach developed below shows, that even for relatively large particles the spectrum is also sparse (see Fig. \ref{fig4} in Sec.~\ref{Method}).

Furthermore, PCM cannot explain the (slightly sloping) shoulder between 1100 and 1250 cm$^{-1}$ in nanodiamond RS  (see, e.g., Ref.~\cite{osswald2009phonon}). To provide an agreement between experimental data and PCM calculations the suggestions of crystal defects and multiple phonon modes have been made, which incorporated even more adjustable parameters into the theory \cite{osswald2009phonon}.

Finally, within the PCM, the vibration mode with bulk optical phonon frequency in the Brillouin zone center $\omega_0=\omega(\textbf{q}=0)$ brings a non-zero contribution to the Raman spectra. Its contribution is suppressed  only due to the low phonon density of states (DOS) in this region, such that lower frequencies (higher momenta) provide more significant contribution to the spectrum. However, the phonon frequency in nanoparticles cannot reach the value $\omega_0$ \textit{in principle}, due to the size quantization effect. 

There were several efforts to improve the PCM by modifying the phonon amplitude envelope in crystallites \cite{zi1997comparison,faraci2006modified,ke2011effect,valentin2008study}, which had only relative success. In Refs. \cite{osswald2009phonon,korepanov2017quantum} the authors accounted for the for phonon dispersion anisotropy in various directions in the Brilloiun zone.

Numerically, the RS of semiconductor and diamond nanoparticles have been modeled by means of the density functional theory (DFT). This approach allowed to calculate the spectra for up to 1000 atoms in a nanocrystallite \cite{zhang2005signature,filik2006raman,li2010convergence}.

In the present paper, we develop a theory free of adjustable parameters except for the phonon line width. We used the material constants unambiguously defined from the microscopic model. The theory is based on the combined use of the dynamical matrix method (DMM) and the bond polarization model\cite{jorioraman} (BPM). It makes possible to describe the RS of nanoparticles with diamond-like lattice. Hereinafter, we shall refer to this theory as the DMM-BPM approach. 

Our method keeps the principal advantage of the PCM intact: the nanoparticle is a zero-dimensional object with respect to the actual wave length of the excitation laser. It allows us to incorporate the bond polarization model in order to treat all the nanoparticle vibration modes.

The primary virtue of the dynamical matrix method is the possibility to study the particles consisting of several thousands of atoms (much larger than what DFT can treat) and to keep the microscopic nature of the description of atomic vibrations. Simultaneously, the classical BPM is famous for successful reproducing of experimental fullerene C60 \cite{snoke1993bond,Guha1996empiricalBPM,menendez2000vibrational} and Si\cite{cheng2002calculations} nanoparticle RS.

More specifically, we utilize the DMM in order to obtain particle vibration modes for nanocrystallites of various shapes and sizes. We derive the bond polarizations using the obtained eigenstates and the material constants known from microscopics. It allows to calculate the nanoparticle RS within the framework of the BPM scheme. Next, from the analysis of relations between the vibrational density of states of a nanoparticle and the calculated spectral intensities, we propose the simple quasicontinuum analytical formulation of the DMM-BPM approach applicable for particles of arbitrary size.

We undertake the thorough comparison of our approach (both analytical and numerical) with the commonly used PCM in order to interpret three detailed sets of experimental data related to nanodiamond RS. We find that our model fits the experiment much better than the PCM one. Note, that we concentrate on nanodiamonds; however, the theory can be easily extended to other crystalline materials.

The rest of the paper is organized as follows. In Sec.~\ref{Theory}, we formulate DMM and BPM in the form suitable for further applications. In order to verify these methods, we supplement our theoretical analysis by \textit{ab initio} calculations within the DFT scheme. In Sec.~\ref{Method}, we combine DMM and BPM into an integrated approach, capable to treat the nanoparticle RS. Furthermore, analyzing the relation between the eigenmodes of nanoparticle and the Raman peak intensity, we formulate the DMM-BPM theory via simple analytical equations. In Sec.~\ref{Experiment}, we present the comparative analysis of our approach and the PCM model, both applied to the available experimental data. In Sec.~\ref{Discussion}, we discuss the advantages of the DMM-BPM and possible extensions of this work. Finally, we present our conclusions. Our paper is supplemented with two Appendixes which contain some details of DMM-BPM scheme.

\section{General theory}
\label{Theory}

\subsection{Dynamical matrix method and bond polarization model}
\label{DMMBPM}

The dynamical matrix method \cite{born1954dynamical,maradudintheory} allows to derive the normal modes and the eigenfrequencies $\omega$ of molecules and nanoparticles by solving $3N \times 3N$ eigenvalue problem
\begin{equation}
  \label{DynMat}
  M\omega^2 u_{i,\alpha}=\sum_{j=1}^N \sum_{\beta={x,y,z}} \frac{\partial^2 \Phi}{\partial   u_{i,\alpha} \partial u_{j,\beta}}u_{j,\beta},
\end{equation}
where $u_{i,\alpha}$ is the $i$-th atom displacement along $\alpha$ direction, $N$ is the number of atoms in the nanoparticle, $M$ is the atomic mass and $\Phi$ is the total energy of the particle as a function of atomic displacements.

We derive $\Phi$ from the microscopic Keating model \cite{keating1966effect,martin1970elastic,kane1985phonon,steiger2011enhanced}. In this model, the parameter $\alpha_0$ measures the bond rigidity with respect to stretching and the parameter $\beta_0$ measures the valence angle bending. Keating model yields the simple  expression for the optical phonon frequency in the $\Gamma$ point:
\begin{equation}
  \omega_0^2=\frac{8}{M}(\alpha_0+\beta_0).
\end{equation}
Substituting $\alpha_0 = 1.068$ Dyn$\cdot$cm$^{-2}$ and $\beta_0 = 0.821$ Dyn$\cdot$cm$^{-2}$ from the Table II of Ref. \cite{anastassakis1990piezo} 

we find $\omega_0=1388$ cm$^{-1}$. This value differs from the one $\omega_0=1333$ cm$^{-1}$ obtained from experimental data\cite{warren1967lattice,schwoerer1998phonon,burkel2001determination,kulda2002overbending} and \textit{ab initio} calculations \cite{monteverde2015under}. In what follows we shall use the renormalized parameters $\alpha=\alpha_0C_N$ and $\beta=\beta_0C_N$, where $C_N=\left(1333/1388\right)^2$. We attribute the rescaling of phonon energies caused by $C_N$ to the regular overestimation of the elastic constants within the Keating model. This procedure provides the correct optical phonon frequency in the Brillouin zone center, keeping the ratio between stretching and bending elastic constants intact.  Note that the formulation of DMM-BPM can be performed using any particular bulk crystal dispersion originating from a specific microscopic model.

To approximate the optical phonon dispersion in diamond we use the standard expression:

\begin{equation}
\label{eq_dispersion}
\omega_{ph}(q)=A+B\cos(\pi \tilde{q}),
\end{equation}
where the normalized $\tilde{q}_{max}=1$ corresponds to the boundary of the Brillouin zone $q_{max}\approx\frac{2\pi}{a_0}$ for diamond-type lattice. Here and below, the quantities with tilde stand for phonon wave vectors normalized to unity; $a_0=0.357$~nm is the diamond lattice constant, and $A+B=\omega_0$.

The Keating model and the employed force constants yield $B\approx 85$ cm$^{-1}$.The bulk diamond phonon dispersion $\omega_{ph}(q)$ calculated on the basis of Keating approach is closer to the one introduced by Ager \cite{ager1991spatially} ($B\approx 91$ cm$^{-1})$ than to those presented by Yoshikawa \cite{yoshikawa1995raman} ($B\approx 141$ cm$^{-1}$) and by Chaigneau \citep{chaigneau2012laser} ($B\approx 32$ cm$^{-1}$).  

The BPM itself and the constants required for deriving the polarization tensors are described in Ref. \cite{jorioraman} (see section 11.5, Eq. (11.16) and Table 11.3) and in Refs. \cite{snoke1993bond,Guha1996empiricalBPM,menendez2000vibrational}. These constants have been introduced for hydrocarbon single bond polarization in Ref. \cite{martin1984raman} and have been adopted for carbon atomic clusters in Ref. \cite{Guha1996empiricalBPM}. The main output of the BPM are the polarization tensors $P_{\alpha\beta}(\nu)$ for $\nu$-th mode. In the most general form, they can be expressed via the normal modes as follows:

\begin{equation}
P_{\alpha\beta}(\nu) = \sum_{i=1}^N \sum_{\alpha'} M_{i,\alpha,\beta,\alpha'} u_{i,\alpha'}(\nu),
\end{equation}
where $M_{i,\alpha,\beta,\alpha'}$ are the combinations of atomic radius vectors and material constants describing the bond polarizations. In Appendix~\ref{AppendixA}, we present the equations expressing the components of tensor $M_{i,\alpha,\beta,\alpha'}$ via the microscopic parameters of the theory.

The quantity $P_{\alpha\beta}(\nu)e_{i\alpha}e_{s\beta}$ characterizes the intensity of photon scattering from the state with polarization $\mathbf{e}_i$ to the state with polarization $\mathbf{e}_s$. In order to calculate the powder RS in the backscattering geometry, one should average the squared vector $P_{\alpha\beta}(\nu)e_{i\alpha}$ over the directions of $\mathbf{e}_s$. We denote the result of this procedure as $\hat{N}(P_{\alpha\beta}(\nu))$. Then, the light intensity of the Raman spectrum $I(\omega)$ can be described as a superposition of Lorentzians centered at corresponding eigenfrequencies with their weights proportional to $\hat{N}(P_{\alpha\beta}(\nu))$,

\begin{equation}
\label{eq_ramanvianormal}
I(\omega)\propto \sum_\nu \frac{n(\omega_{\nu})+1}{\omega_{\nu}}  \hat{N}(P_{\alpha\beta}(\nu)) \frac{\Gamma/2}{(\omega-\omega_\nu)^2+\Gamma^2/4},
\end{equation}
where $n(\omega_{\nu})$ is the Bose-Einstein occupation number for the mode with   frequency $\omega_{\nu}$. The additional parameter of the model is the linewidth $\Gamma$, which is a combination of the spectrometer resolution and the intrinsic phonon damping. In the present paper, we shall treat $\Gamma$ as a free adjustable parameter. Its microscopic origin will be clarified in a separate publication \cite{OurPhononDamping}.

The bond polarization model is closely related to the linear response theory applied previously for studying the RS of glasses\cite{martin1981correlated} and semiconductor nanostructures \cite{elliott1974theory,wang1988theory,alfaro2008theory}. The dynamical matrix diagonalization, calculating RS and further analysis of DOS was performed in ''Mathematica'' package\cite{ram2010}.

\subsection{Ab-initio calculations}
\label{Abinitio}

To verify the usage of dynamic matrix method and bond polarization model for the analysis of RS we perform ab-initio calculations for 0.75 nm, 0.86 nm, 1 nm, and 1.25 nm roughly spherical nanoparticles. The "Gaussian" package \cite{gaussian} was used for calculations with the semi-empirical method PM3\cite{stewart1989optimization} with a standard base 3-21g. The scaling coefficient 0.925 obeys the 1333 cm$^{-1}$ limit value for large particle size when making the approximation of the points obtained by DFT with eq. \eqref{eq_dispersion} and eq. \eqref{eq_omegaL}. The free bonds of the surface atoms were terminated with heavy hydrogens, like in Ref.\cite{filik2006raman}.

\section{Formulating the method}
\label{Method}

\subsection{Basic properties of eigenvalues and eigenfunctions of dynamical matrices}

In our numerics, we consider the nanocrystallites in the shape of a sphere, a cube and a truncated octahedron. Henceforth, we use the diameter $L$ of a spherical particle with the same number of atoms as in the nonspherical ones, as the measure of the effective size of the nonspherical particles. The sizes of particles vary from 0.75 nm to 4.5 nm.  The latter corresponds to approx. 8500 atoms. The eigenvectors of dynamical matrices are normalized to unity and therefore vibration amplitudes are proportional to $L^{-3/2}$, similar to phonon normalization.

For all considered particle shapes and sizes, the structure of the solution of dynamical matrix eigenproblem is as follows. The first three eigenvalues are nearly degenerate and correspond to relative shears of sublattices in three spatial directions. The Raman scattering intensities $\hat{N}(P_{\alpha\beta}(\nu))$ calculated for these eigenfunctions have maximal magnitudes nearly equal to each other. Their sum $N_{1-3}$ is given by

\begin{equation}
N_{1-3}=N_0\frac{L^3}{a_0^3}.
\end{equation}
Here, $N_0$ is the constant calculated within the BPM and corresponding to the set of elastic and polarization parameters we use. It is proportional to the volume of the nanoparticle. Frequencies and Raman intensities of the three highest modes do depend on the particle size, but are almost independent of its shape. On this level of accuracy, the highest eigenfrequency can be found analytically via the formula:

\begin{equation}
\label{eq_omegaL}
\omega_L = \omega_{ph}(q_L),
\end{equation}
where $q_L = \frac{2\pi}{L}$ is the characteristic scale of the momentum size quantization. It is worth mentioning that the main Raman peak is governed by $\omega_L$.

Fig.~\ref{fig1} shows the downshift $\Delta \omega_L=\omega_0-\omega_L$ of the highest mode frequency relative to the bulk diamond one as a function of particle size obtained numerically and its approximation given by Eq. \eqref{eq_omegaL}. Remarkably, $\omega_L$ is almost independent of the shape of a particle. Therefore, the $L$-dependence of $ \omega_L$ in Eq. \eqref{eq_omegaL} allows a simple estimate for the typical nanodiamond size in the powder as a function of the Raman peak position. Moreover, Eqs. \eqref{eq_dispersion} and \eqref{eq_omegaL} permits us to derive simple formulas for RS (see Eqs.\eqref{eq_Nomega}-\eqref{eq_Fq} below).

Similar downshift of the first mode frequency has been reported in Refs. \cite{cheng2002calculations,valentin2008study} for silicon nanoparticles and predicted by toy model in Ref. \cite{meilakhs2016new}. The size effect for the nanoparticle breathing mode has been shown using DFT in Ref. \cite{filik2006raman}.

\begin{figure}
\includegraphics[width=0.45\textwidth]{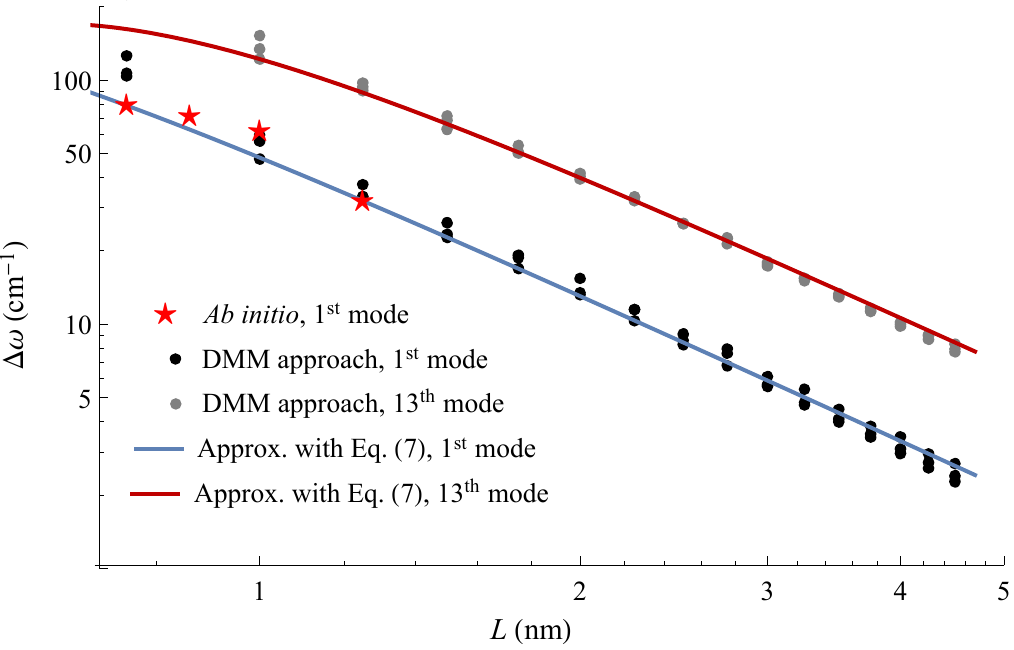}
\caption{\label{fig1} Size dependence of the highest normal mode frequency downshift $\Delta\omega_L=\omega_0 - \omega_L$. Three points for each size (almost indistinguishable) correspond to three nanocrystallite shapes. Black dots are for 1$^{\mathrm{st}}$ mode and gray dots are for 13$^{\mathrm{th}}$ mode. The solid lines drawn for Eq. \eqref{eq_omegaL} fit the numerics well. This fact will be used in our further derivation. The stars show the results of \textit{ab initio} calculations.}
\label{fig1}
\end{figure} 

The origin of these large and almost identical contributions of the first three modes could be understood as follows. Numerically, the magnitudes of atomic displacements are maximal in the centers of the nanoparticles and drop to zero at their boundaries. Generally, they have the cosine-like shape, which is close to the lowest-in-$\lambda$ solution of the continuous equation  $\Delta \psi = \lambda \psi$ with Dirichlet boundary conditions $\psi |_{d\Omega}=0$. \cite{Envelopes}


Numerical analysis (see also Appendix~\ref{AppendixB}) shows that the normal modes from the 4-th to the 12-th are Raman silent, and only the 13-th mode becomes Raman active again. Moreover, the 13-th mode can be treated as the beginning of a band of Raman active modes. This band consists of interleaving Raman active and Raman silent subbands, the subband width being of the order of 10 eigenvalues. The Raman scattering intensity of a single mode within the first subband is at least one order of magnitude weaker than that of the three highest modes. The eigenfrequency corresponding to the 13-th mode can be estimated as $\omega_{L,13} = \omega_{ph}(q_{L,13})$, where $q_{L,13}\approx1.8 q_L$ (cf. Fig. \ref{fig1}). Moreover, the modes starting from the 13-th are very dense. In the next subsection we shall demonstrate that they can be treated as a continuum.

\begin{figure}
\includegraphics[width=0.5\textwidth]{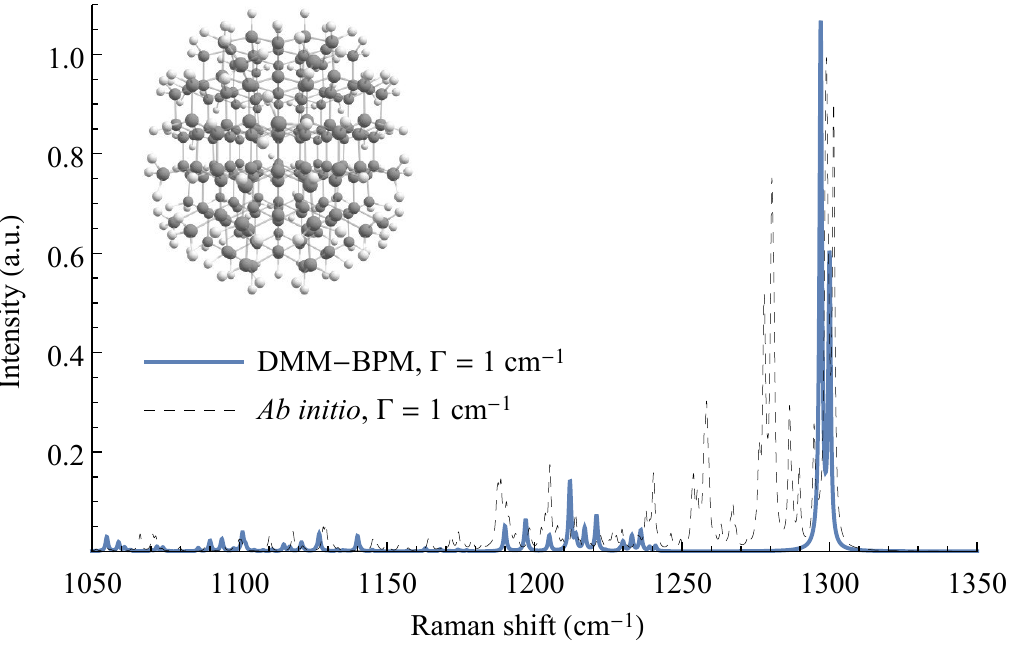}
\caption{Raman spectra of a 1.25 nm spherical nanodiamond particle, obtained by the DMM-BPM approach using Eq. \eqref{eq_ramanvianormal} (blue curve) and \textit{ab initio} (black dashed curve),  with $\Gamma=1$ cm$^{-1}$. The inset shows the 1.25 nm particle with added hydrogens on its surface taken for \textit{ab initio} calculations as in Ref. \cite{filik2006raman}.}
\label{fig2}
\end{figure}

Fig. \ref{fig2} shows the Raman spectra of a 1.25 nm spherical nanodiamond, obtained from Eq. \eqref{eq_ramanvianormal} with $\Gamma=1$ cm$^{-1}$. One can see that the results of DMM-BPM and \textit{ab initio} models yield the close position of the main peak and similar comb-like structure of peaks in the region $1100-1250$ cm$^{-1}$. We postpone the detailed analysis of the nanoparticle shape effect until a forthcoming publication \cite{KFG}.

\subsection{Analytical formulation of DMM-BPM}

Numerically, the dynamical matrix diagonalization and the implementation of BPM become cumbersome for nanoparticles larger than 4 nm. In this subsection, we propose an  approximate analytical scheme for calculating the RS of nanoparticles of arbitrary size based on the analysis of our numerical results. This analysis allows to derive the Raman spectra for larger particles, which cannot be treated within the exact DMM-BPM and/or \textit{ab initio} methods.

\begin{figure}
\includegraphics[width=0.45\textwidth]{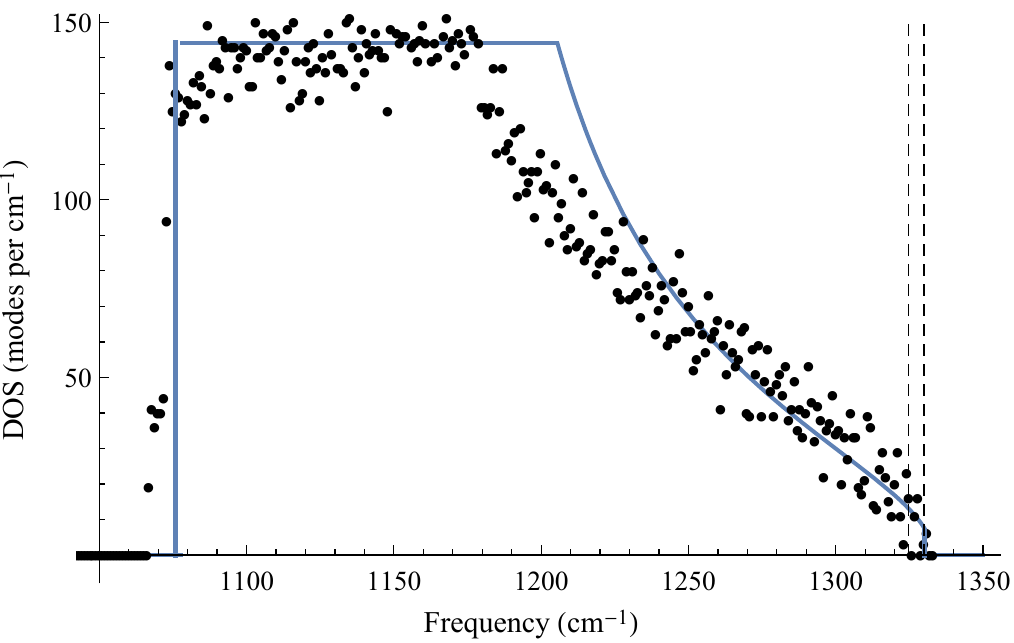}
\caption{\label{figDOS} Sum of phonon DOS of three 4.5 nm diamond nanoparticles (sphere, cube, and truncated octahedron) plotted as a function of number of modes per 1 cm$^{-1}$ (dots). The sum of DOS for three particle shapes is given to enhance the sampling. The blue curve depicts the analytical phonon DOS given by formulas \eqref{eq_DOS} and \eqref{eq_DOSD0}. The two vertical dashed lines show the band of Raman silent modes from $\omega_L$ to $\omega_{L,13}$.}
\label{fig3}
\end{figure}

First, let us approximate the optical phonon DOS as a function of particle size $L$ as follows:

\begin{equation}
\label{eq_DOS}
  D(\omega)=\begin{cases}
    0, & \omega_L < \omega \\
   D_0(\omega), & A-\frac{1}{2}B < \omega < \omega_L \\
   \mathrm{const}=D_0(\omega)|_{\omega=A-\frac{1}{2}B}, & A-2B < \omega < A-\frac{1}{2}B\\
   0, & \omega < A-2B,
  \end{cases}
\end{equation}
where
\begin{multline}
\label{eq_DOSD0}
D_0(\omega)= 4.5 \cdot 10^{-3} \textrm{cm} \cdot \frac{L^3}{a_0^3} \arccos\left( \frac{\omega-A}{B} \right)^2 \\ \times \left[{1-\left( \frac{\omega-A}{B} \right)^2}\right]^{-1/2}.
\end{multline}
Here, the function $D_0(\omega)$ is determined by the bulk DOS derived for the phonon dispersion given by Eq. \eqref{eq_dispersion}. Since the cosine dispersion \eqref{eq_dispersion} fails near the Brillouin zone boundary $\omega=A-B$, and there are several closely lying branches of acoustic and optical modes \cite{monteverde2015under} in this region, we approximate our DOS in the vicinity of the zone boundary by a constant. Eqs.\eqref{eq_DOS} and \eqref{eq_DOSD0} fit well the numerical DOS for diamond nanoparticles (see Fig.~\ref{fig3}), and roughly coincide with the bulk diamond one from Ref. \cite{pavone1993abinitio}, resembling also the Si nanoparticle phonon DOS \cite{valentin2008study,prokofiev2014phonon}.

Dividing numerically the Raman spectrum of nanoparticle by the (numerical) phonon DOS, one can introduce the mean Raman scattering intensity for lower (quasicontinuum) modes $N(\omega)$:

\begin{equation}
\label{eq_NomegaDef}
N(\omega) = \frac{\sum_{\nu:\omega_{\nu}\in (\omega,\omega+d\omega)}\hat{N}(P_{\alpha\beta}(\nu))}{\sum_{\nu:\omega_{\nu}\in (\omega,\omega+d\omega)}1}.
\end{equation}
The numerator of the above equation is the integral Raman intensity for the phonon modes with frequencies lying within the interval $(\omega,\omega+d\omega)$, and the denominator is the total number of such modes.

In spite of discontinuities in scattering intensities of the lower modes in the quasicontinuum and the existence of silent subbands, the energy dependence of the mean Raman scattering intensity can be approximated by a simple power law:
\begin{equation}
\label{eq_Nomega}
N(\omega) = C_1 N_0 \left(1-\frac{\omega}{\omega_L}\right)^{-3/2},
\end{equation}
where $C_1\approx0.003$. This quantity is not proportional to $L^3$, which leads to the overall proportionality of the RS to the nanoparticle volume.

Using Eqs. \eqref{eq_DOS} and \eqref{eq_Nomega}, replacing the summation in Eq.~\eqref{eq_ramanvianormal} by the integration and writing separately the contribution from the three highest modes, we obtain

\begin{figure}[t]
\includegraphics[width=0.5\textwidth]{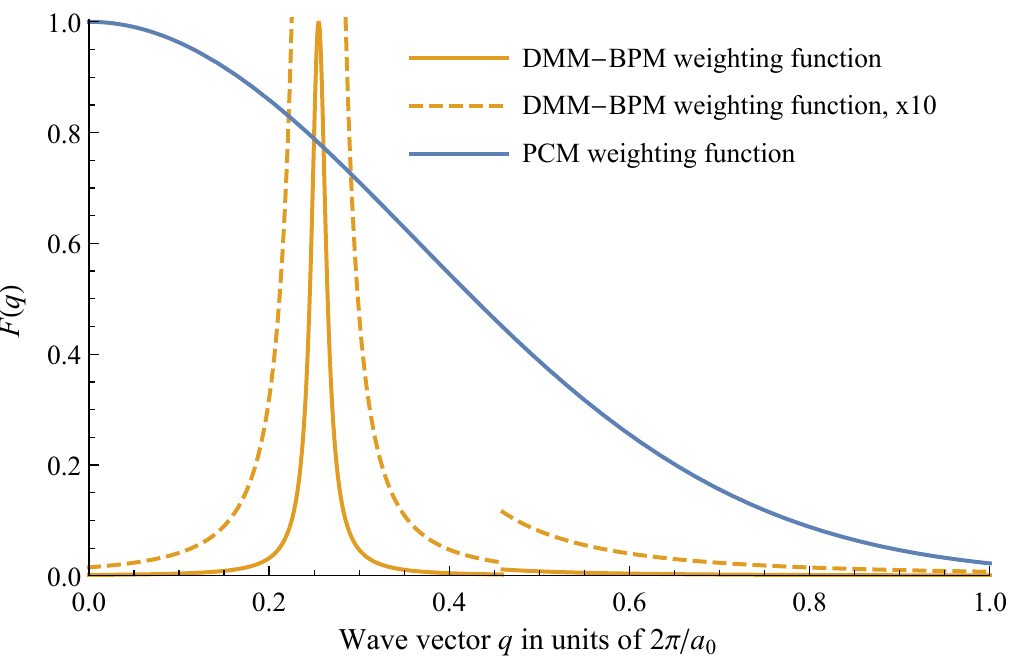}
\caption{Function $F(\tilde{q})$ to be integrated to yield the Raman spectrum, see Eq.~\ref{eq_Iq}. The size of the particle is 1.4 nm. The yellow solid curve is for present theory Eq. \eqref{eq_Fq}, and the blue curve is for the standard PCM envelope $F(\tilde{q})=\exp (- \tilde{q}^2L^2/4a_0^2)$. The yellow dashed curve is for $F(\tilde{q})$ given by present theory and multiplied by the factor of 10. Note the discontinuity due to the Heaviside function. }
\label{figFq}
\end{figure}

\begin{multline}
\label{eq_Iomega}
I_L(\omega) = \frac{N_0L^3}{a_0^3}\frac{1}{\pi}\frac{\Gamma/2}{(\omega-\omega_L)^2+\Gamma^2/4} \\ +\int_0^{\omega_{L,13}} \frac{d\omega^\prime}{\pi}\frac{D(\omega^\prime)N(\omega^\prime) \, \Gamma/2}{(\omega-\omega')^2+\Gamma^2/4}.\end{multline}

The first term in this equation related to the three highest modes is the dominating contribution to the Raman peak. It cannot be described within the continuum approach. Note also that the upper integration limit in the ``continuum term'' is $\omega_{L,13}$. One can see that both terms (and therefore the entire spectrum $I_L(\omega)$) are proportional to $N_0 L^3$. Eq. \eqref{eq_Iomega} is close by its meaning to Eq. (13) from Ref. \cite{nemanich1981light} and clarifies the physical meaning of the coupling constant $C$ there.

It is instructive to compare this analytical approach and the PCM. With a little loss of accuracy (namely, neglecting the contribution of phonons with frequencies below $A-B$ and omitting some fine structure of nanoparticle vibrations with frequencies close to $\omega_{ph}(q_L)$), one can rewrite Eq. \eqref{eq_Iomega} as a PCM-like integral:

\begin{equation}
\label{eq_Iq}
I_L(\omega)\propto L^3 \int_0^1 \frac{\tilde{q}^2d\tilde{q} F(\tilde{q}) \, \Gamma/2}{(\omega-\omega_{ph}(\tilde{q}))^2+\Gamma^2/4},
\end{equation}
with $\omega_{ph}(q)$ being the standard bulk diamond dispersion, and
\begin{equation}
\label{eq_Fq}
F(q)=\delta(\tilde{q}-\tilde{q}_L)/\tilde{q}_L^2+C_2N(\tilde{q})\theta(\tilde{q}-\tilde{q}_{L,13}),
\end{equation} 
where $N(\tilde{q})=(\tilde{q}^2 - \tilde{q}_L^2)^{-3/2}$ is the averaged scattering intensity in the momentum domain, $\theta(x)$ is the Heaviside function, and $C_2 \approx 0.2$. The contribution from the highest modes (delta-function term) enters Eq.~\eqref{eq_Fq} separately from the phonon continuum (theta-function term). A comparative analysis of the kernel $F(q)$ given by Eq.~\ref{eq_Fq} and by PCM is presented in Fig.~\ref{figFq}.  

In our approach, the function $F(q)$ plays the role of the averaged Raman scattering intensity for phonons with wave vector $q$. On the contrary, in the PCM approach, this function $F(q)$ is assumed to be the convolution of the phonon amplitude envelope Fourier image. Furthermore, in our theory the two-component function $F(q)$ is significantly sharper than the Gaussian function used in the PCM, due to the assumption of hard boundaries of particles (see Fig.~\ref{figFq}).

Formulas \eqref{eq_Nomega} and \eqref{eq_Iomega} constitute our main result in $\omega$-representation, whereas formulas \eqref{eq_Iq} and \eqref{eq_Fq} are the same in $q$-representation (PCM-like). They can be directly utilized for calculation of Raman spectra of arbitrary-sized nanodiamonds.

\begin{figure}
\includegraphics[width=0.5\textwidth]{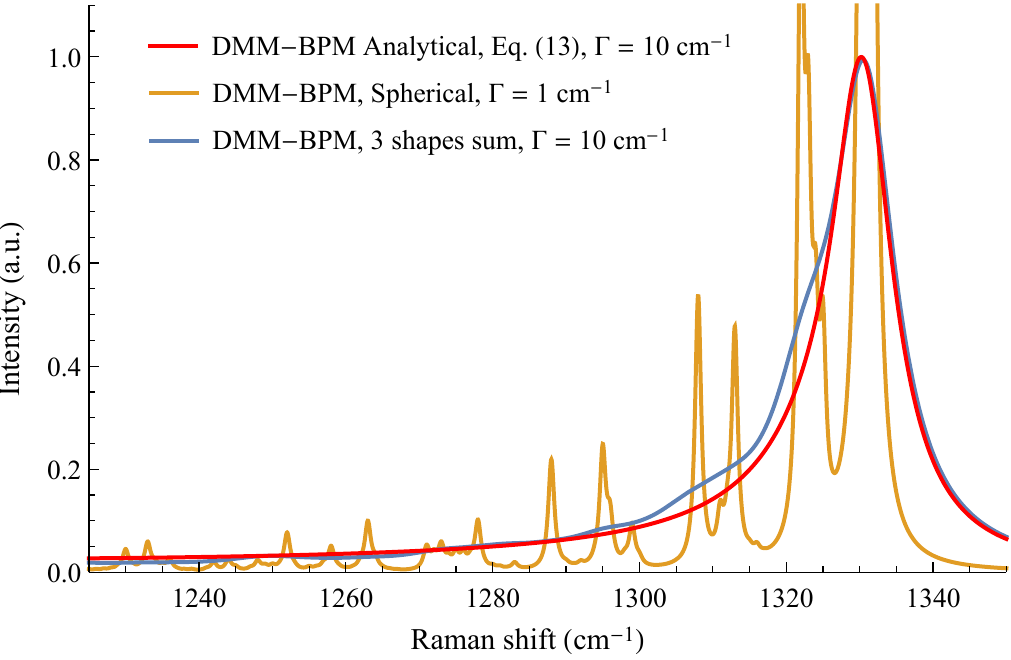}
\caption{Raman spectra of 4.5 nm spherical nanodiamond particle obtained within the exact DMM-BPM with $\Gamma=1$ cm$^{-1}$ (orange curve), the average of three Raman spectra of 4.5 nm particles (sphere, cube, and truncated octahedron) obtained within the DMM-BPM  with $\Gamma=10$ cm$^{-1}$ (blue curve) and the spectrum obtained analytically using Eq. \eqref{eq_Iq} with $\Gamma=10$ cm$^{-1}$ (red curve). The height of the orange peak is 9.7. It is cut for clarity.}
\label{fig4}
\end{figure}

Fig. \ref{fig4} shows the Raman spectra of a 4.5 nm spherical nanodiamond obtained with Eq. \eqref{eq_ramanvianormal} and $\Gamma=1$ cm$^{-1}$: the sum of three RS for 4.5 nm particles (sphere, cube, and truncated octahedron) with $\Gamma=10$ cm$^{-1}$ and the spectrum obtained using Eq.~\eqref{eq_Iq} with $\Gamma=10$ cm$^{-1}$. First, we see the pronounced comb-like structure of the spectrum for smallest $\Gamma$ even for substantially large 4.5 nm particles. Second, taking various nanoparticle shapes and introducing broader $\Gamma$ result in effective averaging of the above-mentioned spectrum. Third, we see that the analytical formula Eq.~\eqref{eq_Iq} reproduces very well the result of exact DMM-BPM. Therefore, Fig.~\ref{fig4} demonstrates that our analytical approach is very useful for the treatment of mixtures of differently shaped nanoparticles. 

To conclude this subsection, we note that for nanoparticle powder with a broad size distribution $n(L)$ one should derive the RS via summation over particle sizes:

\begin{equation}
  \label{IWL}
I(\omega)=\sum_{L}I_L(\omega)n(L).
\end{equation} 

\section{Comparing with experiment}
\label{Experiment}

In order to verify our theory, let us consider recent scrupulous experimental measurements of nanodiamond RS accompanied by the analysis of particle size distributions \cite{stehlik2015size,stehlik2016high,shenderova2011nitrogen}.

\begin{figure}
\includegraphics[width=0.5\textwidth]{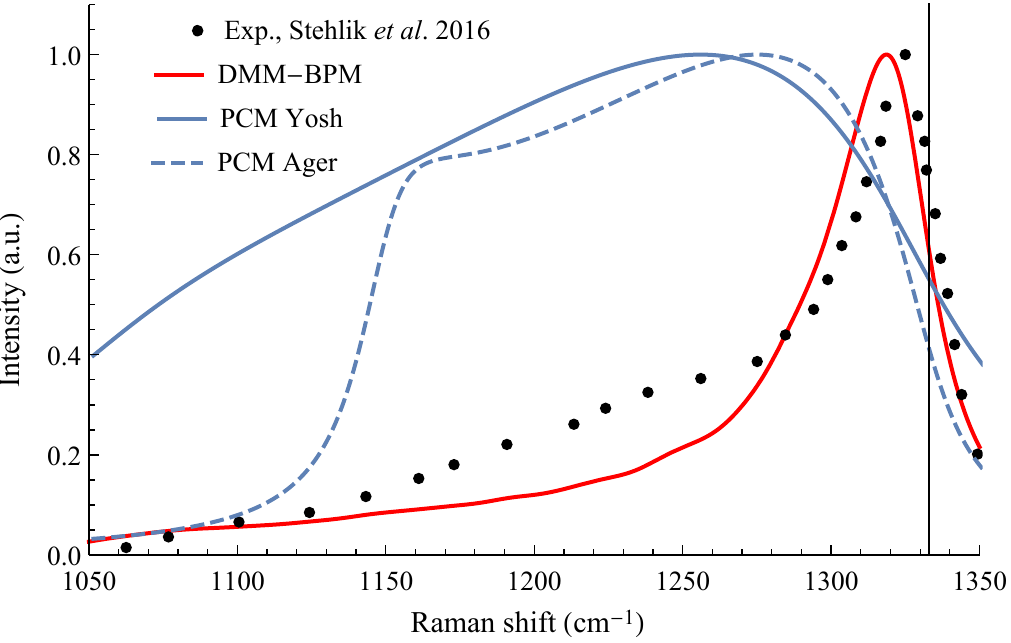}
\caption{Comparison of the experimental Raman spectra from Ref. \cite{stehlik2016high} for nanodiamonds ranged between 0.6 nm and 3.1 nm (black dots), predictions of the PCM (blue curves) and the present theory (red curve) based on the nanoparticle distribution from Ref. \cite{stehlik2016high}. Vertical black line denotes the position of the bulk diamond peak.}
\label{fig5}
\end{figure}

We calculate numerically the RS of nanoparticles within the framework of the exact DMM-BPM using the size distribution taken from Fig. 1(e) of Ref. \cite{stehlik2016high} (red curve in our Fig.~\ref{fig5}). This distribution contains particles from 0.6 nm to 3.1 nm with  a maximum around 1.4 nm; for each particle size, we consider three geometric shapes (spheres, cubes and truncated octahedra). Further, in Fig.~\ref{fig5} we demonstrate the original experimental RS (black dots) from Ref.~\cite{stehlik2016high} with the peak position at 1324 cm$^{-1}$ as well as the fit of these data that utilizes the PCM. The correctness of the PCM parameters used here is justified by the complete reproduction of the RS plots within the Yoshikawa (Fig.~8 of Ref.~\cite{stehlik2015size}, $L=10.8$ nm) and Ager (Fig.~2(a) of Ref. \cite{osswald2009phonon}) models, respectively.

From Fig.~\ref{fig5}, we see a very good agreement between our calculations and the experimental data of Ref.~\cite{stehlik2016high} (except for the underestimation of the shoulder magnitude).  The PCM spectra are drastically different from both ours and Ref.\cite{stehlik2016high} results; they are unable to fit the experimental data for small nanoparticles. Furthermore, our peak position (1320 cm$^{-1}$) is much closer to the experimental value (1324 cm$^{-1}$) than that of the PCM (1260 - 1300 cm$^{-1}$). According to Fig.~\ref{fig1}, the peak downshift equal to 13 cm$^{-1}$  corresponds to 2 nm nanoparticles, which matches the mean value of the size distribution. We use $\Gamma=28 $cm$^{-1}$ for DMM-BPM and PCM with Ager dispersion, and $\Gamma (L) = (3.0+\frac{145  \mathrm{nm}}{L}) \cdot 1$ cm$^{-1}$ for PCM with Yoshikawa dispersion (see also Eq.~(4) in Ref.~\cite{osswald2009phonon}).

\begin{figure}
\includegraphics[width=0.5\textwidth]{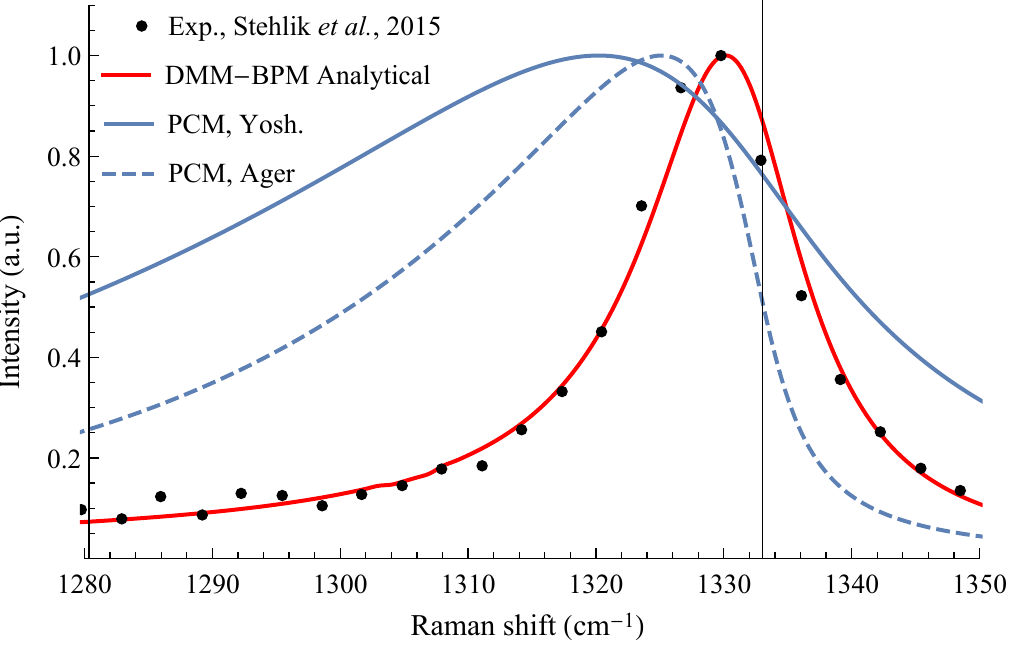}
\caption{Comparison of experimental Raman spectra from Ref. \cite{stehlik2015size} (black dots) for nanodiamonds ranged between 3 and 10 nm, predictions of PCM (blue curves) and the present theory (red curve) based on the nanoparticle distribution from Ref. \cite{stehlik2015size}. Vertical black line denotes the position of the bulk diamond peak.}
\label{fig6}
\end{figure}

Next, in the experiment of Ref. \cite{stehlik2015size}, the size distribution contains larger nanoparticles (up to 10 nm). Therefore, it is tempting to employ the analytical approach of our Eq.~\eqref{eq_Iq}. It is depicted in Fig.~\ref{fig6} by the red curve. The MSY18-O1 sample Raman spectrum from Fig.~8 of Ref.~\cite{stehlik2015size} is plotted in Fig.~\ref{fig6} using black dots, together with PCM realization with Ager and Yoshikawa dispersions. The width $\Gamma=28$ cm $^{-1}$ is taken for analytical DMM-BPM approach and PCM with Ager dispersion. For Yoshikawa dispersion, $\Gamma(L)$ is taken as previously.



Fig. \ref{fig6} demonstrates an excellent agreement between our theory in analytical form and the experiment. In particular, there are no problems with the shoulder description as it occurred for smaller particles. The PCM is working better than for smaller particles, but the quality of the fit is still far from being acceptable. Experimental peak position is 1329.6 cm$^{-1}$, while the developed theory predicts 1330.3 cm$^{-1}$.

We arrive to similar conclusions analyzing the data of Ref.~\cite{shenderova2011nitrogen}. Again, DMM-BPM is applied on the analytical level of Eq.~\eqref{eq_Iq}. We examine the experimental Raman spectrum of ND-TNT/RDXd sample. The authors of Ref.~\cite{shenderova2011nitrogen} point out that XRD gives 4 nm for mean nanoparticle size and HRTEM yields 3-6 nm for width of the distribution; however, the entire histogram is not presented. In order to mimic the data of Ref.~\cite{shenderova2011nitrogen} we artificially construct a normal distribution with FWHM = 1 nm centered at 4 nm. We choose $\Gamma=32 $ cm$^{-1}$ for analytical DMM-BPM approach and PCM with Ager dispersion. For PCM with Yoshikawa dispersion, $\Gamma(L)$ is taken as previously.

\begin{figure}
\includegraphics[width=0.5\textwidth]{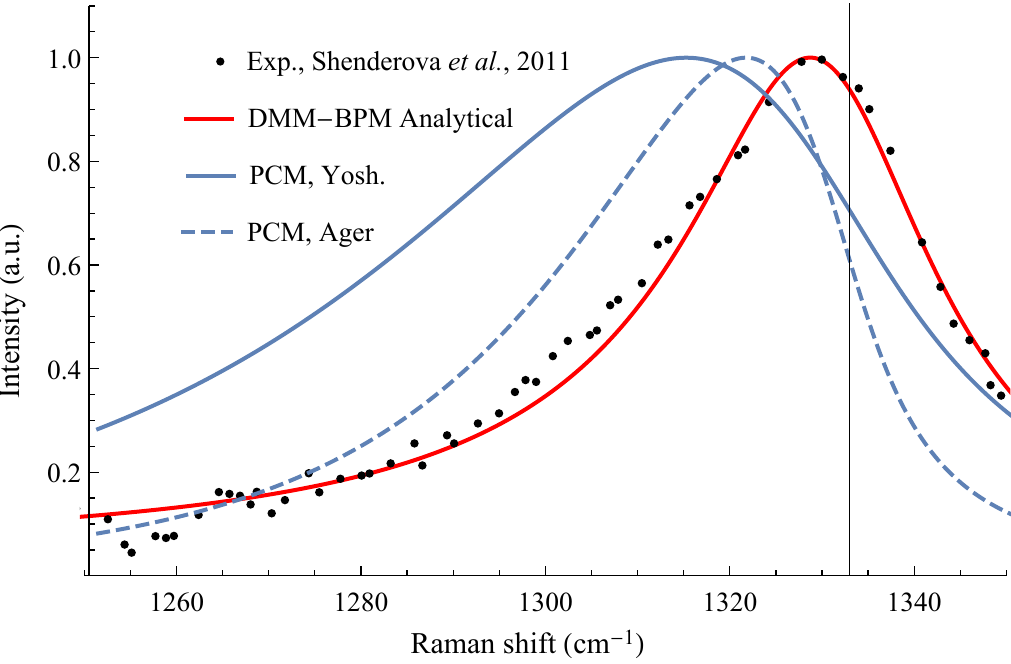}
\caption{Comparison of an experimental Raman spectrum from Ref. \cite{shenderova2011nitrogen} (black dots), predictions of PCM for nanodiamonds of approximately 4 nm (blue curve) and the present DMM-BPM theory based on a Gaussian distribution centered at 4 nm (red curve) are shown. Vertical black line denotes the position of the bulk diamond peak.}
\label{fig7}
\end{figure}

The impressive success of DMM-BPM theory as compared to the PCM in fitting the experimental data of Ref.~\cite{shenderova2011nitrogen} is drawn in Fig.~\ref{fig7}.

It worth mentioning, that the use of Ager phonon dispersion with lower $B$ coefficient leads to the Raman peak downshift approximately 1.5 times smaller with respect to Yoskikawa dispersion.

The analysis of this subsection clearly demonstrates the advantages of our theory in the interpretation of the experimental data, especially in the range of small nanoparticles, where the PCM evidently fails. 

\section{Discussion and conclusion}
\label{Discussion}

\subsection{Discussion}

The developed exact DMM-BPM theory is based on a microscopic description of vibrational properties of the system and is free of nonphysical adjustable parameters. The fitting parameter $\Gamma$, which we use in our calculations, has a clear physical meaning of a  phonon linewidth. It needs further microscopic clarification \cite{OurPhononDamping}. 

The DMM-BPM provides the solid ground for a very simple analytical version. Moreover, the exact DMM-BPM is especially useful for the treatment of small nanoparticles, for which the PCM is not applicable and the only competitor of our method is the completely numerical \textit{ab-initio} methods.

Furthermore, the developed analytical formulation of the DMM-BPM theory explains the shapes of nanodiamond RS and their key features from a new point of view. According to this view, the spectra are the superpositions of two contributions. The first contribution is due to the highest nanoparticle vibration eigenmodes, and the second one, stemming from a quasicontinuum, provides the structure of a shoulder of RS.

\begin{figure}
\includegraphics[width=0.5\textwidth]{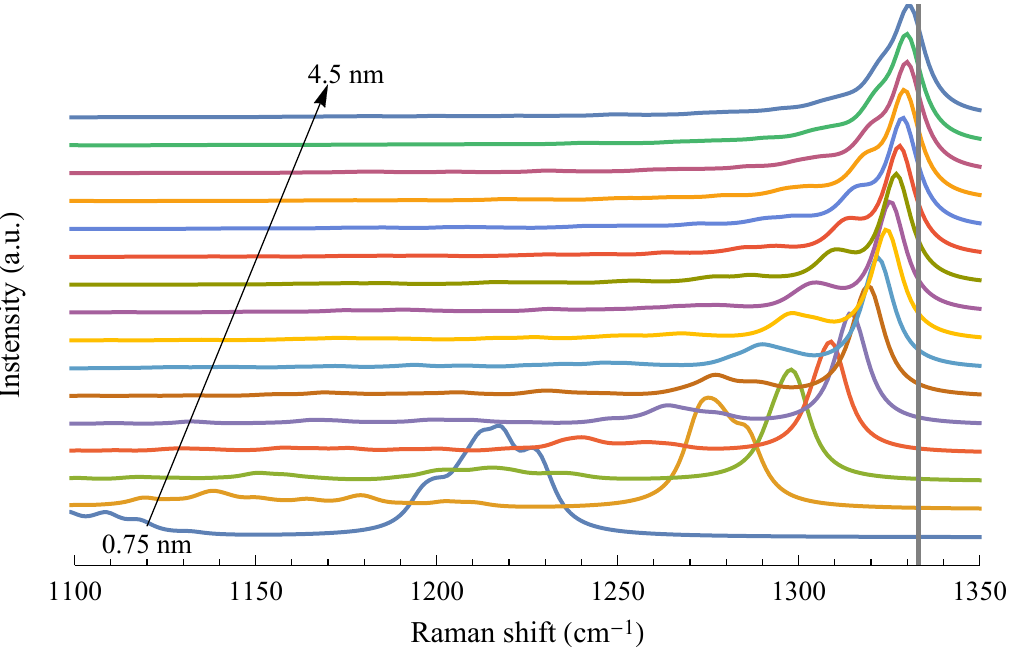}
\caption{Raman spectra of diamond nanoparticles from 0.75 nm to 4.5 nm in size obtained numerically using the exact DMM-BPM approach. $\Gamma=10$ cm$^{-1}$. With increasing of the particle size we observe (i) the smearing of the fine peak structure, (ii) the sharpening of the peak, and (iii) its tendency to the vertical black line which denotes the position of the bulk diamond peak.}
\label{fig_spectraall}
\end{figure}

Due to its dispersion, the optical phonon with the maximal energy has a maximal wavelength, which is of the order of the nanoparticle size. This unambiguously connects the main peak position with the size of the particle, see Fig.~\ref{fig1}. With decreasing the size, the phonon energy decreases and the main Raman peak becomes sharper and shifts down. This process is schematically depicted in Fig.~\ref{fig_spectraall}.

Despite the fact that a single crystallite RS consists of multiple narrow peaks, averaging over particle shapes and sizes as well as taking into account the intrinsic phonon damping leads to smoothing of spectra for real powders, which can be seen from comparison of theoretical spectra and experimental data.

The DMM-BPM reproduces the experimental Raman spectra obtained recently with a very high accuracy. However, it requires the information about the size distribution function. On the contrary, the PCM is not able to reproduce these spectra (especially for small particles) and contains critical imperfections \cite{zi1997comparison}. The developed model can be also formulated as an integral over the Brillouin zone, thus resembling the PCM with different in physical meaning and shape (non-Gaussian) and well-motivated weight function $F(q)$. It makes the application of our theory simple and straightforward. The DMM-BPM provides the Raman spectroscopy with a novel useful tool for nanoparticle size measurements along with DLS, XRD, HRTEM, SAXS, and AFM.

In Refs. \citep{yoshikawa1995raman,aleksenskii1997diamond,osswald2009phonon,chaigneau2012laser} the experimental Raman spectra of nanodiamonds have been presented, providing the mean sizes of nanoparticles. However, the information about the size distribution function is not available. We believe that our theory should be successful in interpreting these experimental results \citep{yoshikawa1995raman,aleksenskii1997diamond,osswald2009phonon,chaigneau2012laser}, given that more detailed information will be at hands.

Meanwhile, there are some experiments that cannot be interpreted directly within the framework of the developed approach. For instance, Ref. \cite{yoshikawa1995raman} reports a 10 cm$^{-1}$ downshift for 4.3 nm nanodiamonds, measured by XRD, while our model yields a twice smaller value. This phenomenon can be attributed to the effect discussed in Ref. \cite{chaigneau2012laser}, where the Raman peak downshift 3.5 cm$^{-1}$ has been reported for nanodiamonds of 9 nm  size measured by DLS. The authors of Ref. \cite{chaigneau2012laser} argued from HRTEM data that these 9 nm particles are actually the aggregates of two or three primary units, giving 3-6 nm for their sizes. However, this speculation is out of scope of our study.

Finally, the distorted and partly amorphous diamond surface can contribute to the shoulder in RS between 1100 cm$^{-1}$ to 1250 cm$^{-1}$ \cite{ferrari2004raman,korepanov2017carbon}, because there are evidences of its changes upon nanoparticle oxidation \cite{stehlik2015size,osswald2009phonon}. The amorphous phase with an intermediate sp$^{3-x}$ hybridization on the nanodiamond surface can also contribute to the Raman signal as sp$^2$ band at 1600 cm$^{-1}$. Accounting for the nanoparticle surface reconstruction and amorphization is also beyond the scope of the present work.

\subsection{Conclusion}

We propose a novel theory of Raman scattering in crystalline nanoparticles, free of adjustable parameters and unphysical assumptions. This theory consists of two ingredients, namely the dynamical matrix method and the bond polarization model (DMM-BPM approach). The theory allows a simple analytical formulation for large enough ($\geq 2$ nm in diameter) particles. We calculate the nanodiamond Raman spectra and find very good to excellent agreement between our theory and the available experimental data, especially in the range of parameters where the commonly used phonon confinement model (PCM) fails.

\begin{acknowledgments}
The contribution to the study conducted by S.V.K. was funded by RFBR according to the research project 18-32-00069. We acknowledge the support of the project "Quantum Fluids of Light" (ANR-16-CE30-0021). O.I.U. thanks for financial support the Skolkovo Foundation (grant agreement for Russian educational and scientific organisation no.7 dd. 19.12.2017) and Skolkovo Institute of Science and Technology (General agreement no. 3663-MRA dd. 25.12.2017). A.V.S. acknowledges the computational grant G60-8 from the Interdisciplinary Centre for Mathematical and Computational Modeling (ICM) in Warsaw. We are gratefully indebted to Guillaume Malpuech for fruitful discussions and attention to work.
\end{acknowledgments}

\appendix

\section{Polarization tensor in BPM}
\label{AppendixA}

Due to the main assumption of the empirical BPM, the nanoparticle electronic polarizability can be expressed as a sum of individual bond polarizabilities. Furthermore, the bond polarizability is assumed to be the function of bond length $R$ and does not depend on chemical environment. The explicit formula includes isotropic and anisotropic parts,

\begin{equation}
  \label{Pol1}
  P_{\alpha\beta} = \frac{1}{3} (\alpha_\parallel+2 \alpha_\perp) \delta_{\alpha \beta} +    (\alpha_\parallel- \alpha_\perp) \left( \frac{R_\alpha R_\beta}{R^2} - \frac{1}{3}   \delta_{\alpha \beta} \right), 
\end{equation}
where $\alpha$ and $\beta$ are the Cartesian coordinates, $\alpha_\parallel$ and $\alpha_\perp$ are the polarizability parameters. The intensity of the first-order Raman scattering reads:
\begin{equation}
  I_{is}(\omega) \propto \sum_\nu \frac{n(\omega_{\nu})+1}{\omega_{\nu}} \left| \sum_{\alpha \beta } P_{\alpha\beta}(\nu) e_{i\alpha} e_{s\beta} \right|^2\delta(\omega - \omega_f).
\end{equation}
Here, $\mathbf{e}_i$ and $\mathbf{e}_s$ are the polarizations of incident and scattered light, respectively, $n(\omega_{\nu})$ is the Bose-Einstein occupation number for $\nu$-th phonon mode with frequency $\omega_\nu$. The tensor $P_{\alpha \beta}(\nu)$ can be expressed via the electronic polarizability derivatives with respect to the real space atomic displacements $u_{i,\alpha}$ as follows:
\begin{equation}
  \label{Pol2}
  P_{\alpha \beta}(\nu) = \sum_{i, \gamma} \frac{\partial P_{\alpha\beta}}{\partial u_{i,\gamma}} u_{i,\gamma}(\nu),
\end{equation}
where $i$ enumerates all the atoms in a nanoparticle and $u(\nu)$ is the eigenvector of the dynamical matrix eigenproblem \eqref{DynMat}. After some calculations, Eqs.~\eqref{Pol1} and \eqref{Pol2} yield
\begin{eqnarray}
   P_{\alpha \beta}(\nu) = &-& \sum_{i,B} \Biggl[ \mathbf{e}_0(i,B) \cdot \mathbf{u}_i(\nu) \Biggl\{\left( \frac{\alpha^\prime_\parallel(B)+2 \alpha^\prime_\perp(B)}{3}\right) \delta_{\alpha \beta} \nonumber \\ &+& (\alpha^\prime_\parallel(B)- \alpha^\prime_\perp(B)) \left( e_{0 \alpha}(i,B) e_{o \beta}(i,B)- \frac{1}{3} \delta_{\alpha \beta} \right)\Biggr\} \nonumber \\ &+& \left( \frac{\alpha_\parallel(B)- \alpha_\perp(B)}{R_0(i,B)} \right) \{ e_{0 \alpha} u_{i,\beta}(\nu) - e_{0 \beta} u_{i,\alpha}(\nu) \nonumber \\ &-& 2\mathbf{e}_0(i,B) \cdot \mathbf{u}_i(\nu) e_{0 \alpha}(i,B) e_{o \beta}(i,B) \}  \Biggr], 
\end{eqnarray}
where $B$ denote the bonds of the $i$-th atom, $\mathbf{e}_0(i,B)=\mathbf{R}_0(i,B)/R_0(i,B)$ is the unit vector along the bond $B$, and $\alpha^\prime_\parallel(B)$ and $\alpha^\prime_\perp(B)$ are the radial derivatives of polarizability parameters.

\section{Active and silent modes}
\label{AppendixB}

\begin{figure}[b]
\includegraphics[width=0.5\textwidth]{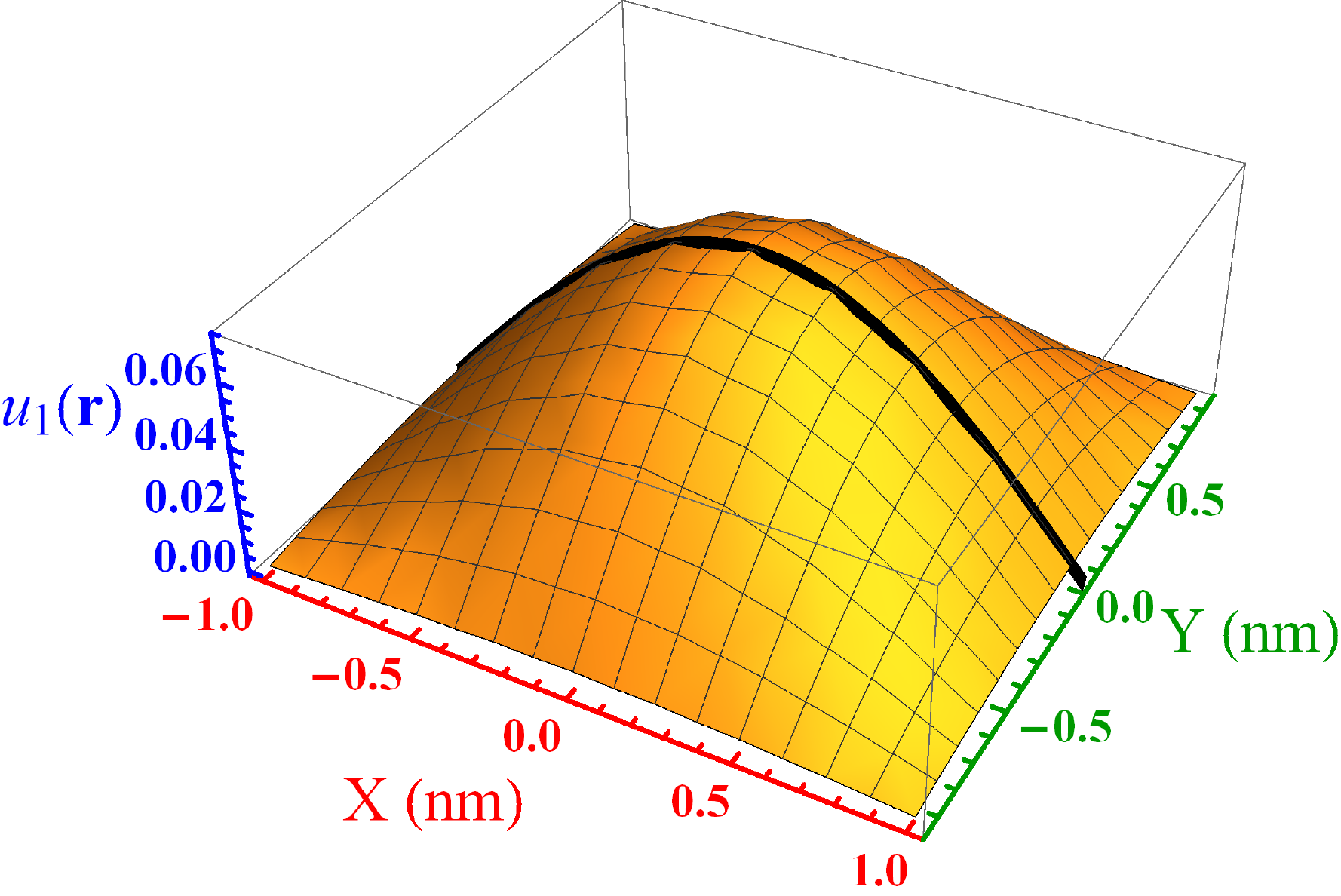}\\
\includegraphics[width=0.5\textwidth]{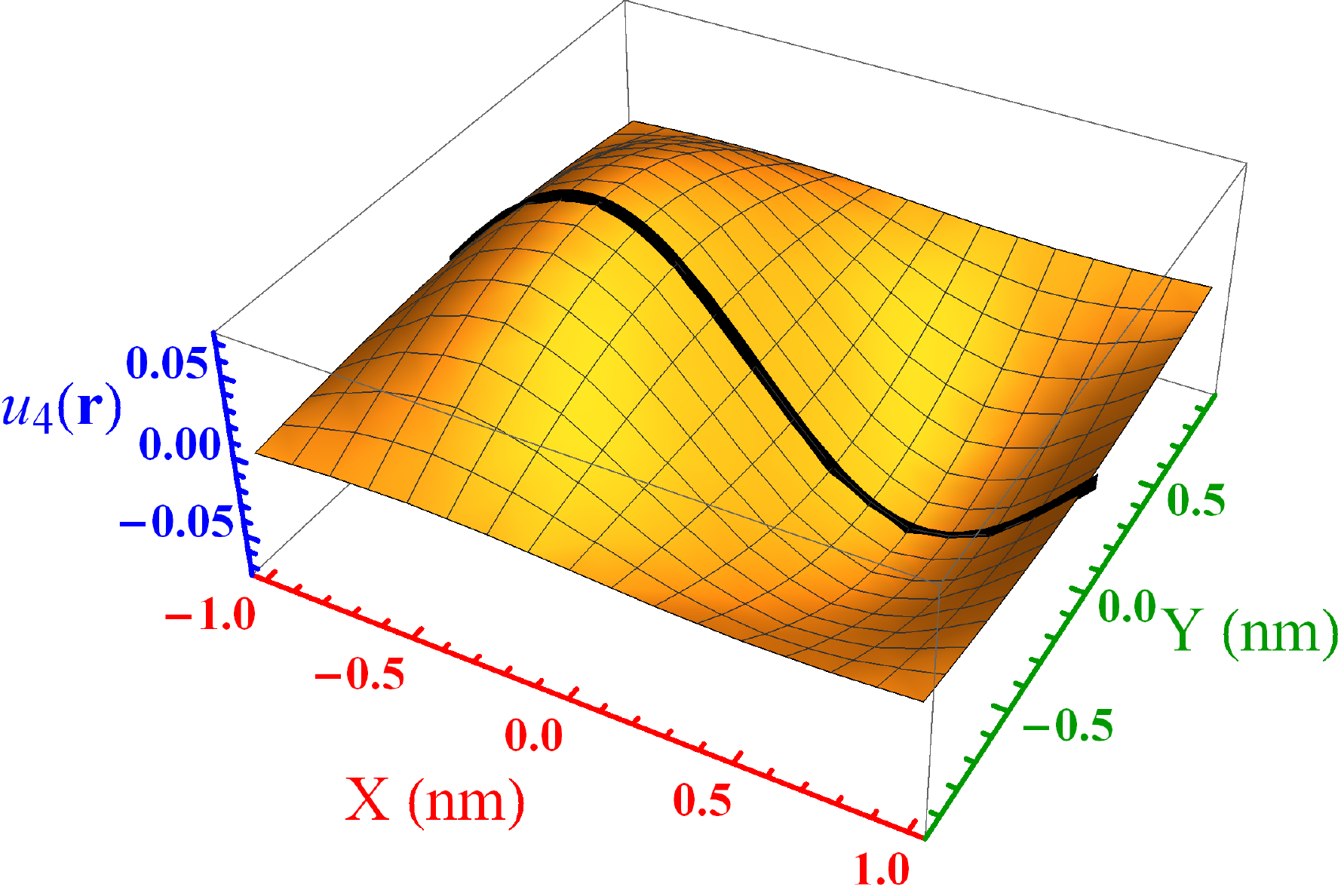}
\caption{Spatial dependence of the phonon eigenfunctions $u_1(\mathbf{r})$ (upper plot) and $u_4(\mathbf{r})$ (lower plot). The integral of the normalized eigenfunction   over the nanoparticle volume is of order of unity for $u_1(\mathbf{r})$ and tends to zero for $u_4(\mathbf{r})$. Atoms are shown only with an absolute value of Z coordinate $<0.15$ nm giving the XY plane cross-section of the nanoparticle. Red and green axes are for X and Y atomic positions, respectively, and blue axes are for $u_1$ and $u_4$ amplitudes. Displacements for two sublattices are taken with opposite signs due to the nature of optical phonons. Black curve is the sine function along X axis.}
\label{figSilent}
\end{figure}

Within the BPM (See Appendix \ref{AppendixA}) the polarizability tensor $P_{\alpha \beta}(\nu)$ is the superposition of individual bond polarizabilities summed up over the nanoparticle volume. The latter ones are determined by the phonon wavefunctions.  In Fig.~\ref{figSilent} (upper panel), we plot the spatial structure of the 1$^\mathrm{st}$ eigenmode of a 2.5 nm cubic nanoparticle as an example of a ``Raman active'' mode. It can be approximated by the envelope function $u_1(\mathbf{r}) = \cos(\pi r_x/L)\cos(\pi r_y/L)\cos(\pi r_z/L)$, which is evidently positive, thus its integral over the nanoparticle volume yields a large positive contribution to the Raman intensity. On the other hand, the 4$^\mathrm{th}$ eigenmode depicted in Fig.~\ref{figSilent}(b) is an example of a ``Raman silent'' mode, as it changes its sign across the crystal and has a node near the crystal center. With a good accuracy, it can be approximated by $u_4(\mathbf{r}) = \sin(2\pi r_x/L)\cos(\pi r_y/L)\cos(\pi r_z/L)$, so its spatial integral tends to zero. Similar analysis can be performed for any other mode. This justifies our segregation of phonon eigenmodes into Raman active and Raman silent parts.

\bibliography{raman}

\end{document}